\documentclass[prb,twocolumn,aps,showpacs]{revtex4}
\usepackage{graphicx}
\usepackage{amsmath}
\usepackage{subfigure}
\usepackage{epstopdf}
\begin{document}
\newcommand{\s}{\scriptscriptstyle}
\newcommand{\uu}{\s \uparrow \uparrow}

\newcommand{\du}{\s \downarrow \uparrow}
\newcommand{\ud}{\s \uparrow \downarrow}
\newcommand{\dd}{\s \downarrow \downarrow}

\title{Analytical study of spin-dependent transition rates within pairs of dipolar and strongly exchange coupled spins with ($S=1/2$) during magnetic resonant excitation}
\author{R. Glenn, M. E. Limes, B. Saam, C. Boehme,  and M. E. Raikh}
\affiliation{Department of Physics and Astronomy, University of Utah, Salt Lake City, UT 84112}
\begin{abstract}
We study theoretically the spectrum, ${\bf\mathrm{F}}(s)$, of spin--dependent transition rates within dipolar $D$ and exchange $J$ coupled pairs of two spins with $S=1/2$ undergoing Rabi oscillations due to a coherent magnetic resonant excitation. We show that the Rabi oscillation controlled rates exhibit a spectrum with three frequency components. When exchange is stronger than the Rabi frequency ($J\gg\Omega_{\s R}$), the frequency components of the Rabi oscillation do not depend on $J$, rather they are determined by the relation between $\Omega_{\s R}$ and $D$. We derive analytical expressions for the frequencies and the intensities of all three Rabi oscillation components as functions of $\Omega_{\s R}/D$ and $\delta/D$, where $\delta$ is detuning of the driving ac field from the Larmor frequency. When $\Omega_{\s R} \gg D$, the two lower frequencies approach $s=\Omega_{\s R}$, while the upper line approaches $s=2\Omega_{\s R}$. Disorder of the local Larmor frequencies leads to a Gaussian broadening of the spectral lines. We calculate corresponding widths for different $\Omega_{\s R}/D$ and $\delta/D$. Unexpectedly, we find that one of the frequency components exhibits an unusual evolution with $\Omega_{\s R}$: its frequency {\em decreases} with $\Omega_{\s R}$ at $\Omega_{\s R}<D$. Upon further increase of $\Omega_{\s R}$ this frequency then passes through a minimum and, eventually, approaches $s=\Omega_{\s R}$. Nonmonotonic behavior of the frequencies is accompanied by nonmonotonic behavior of the respective oscillation intensity.
\end{abstract}
\maketitle

\section{Introduction}

\begin{figure}[t]
\begin{center}
\includegraphics[width=90mm]{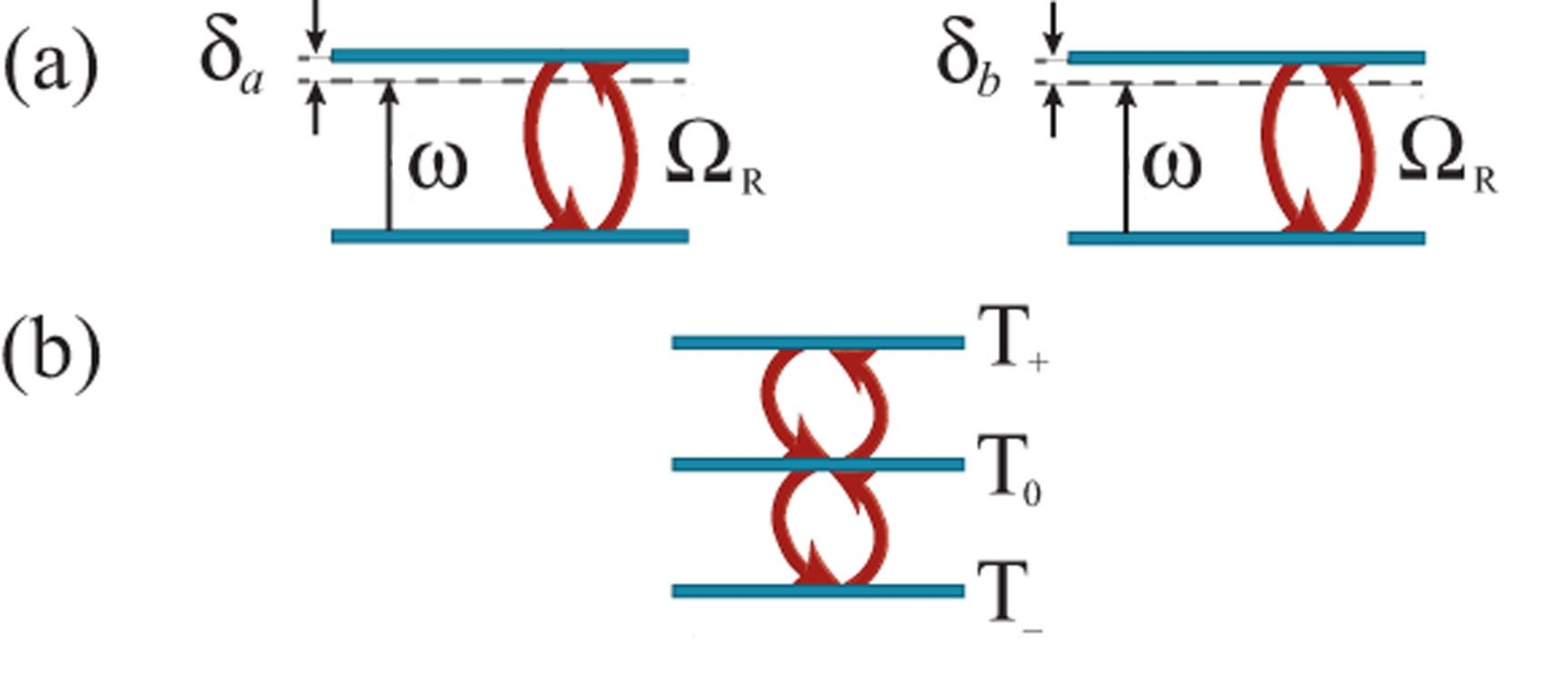}
\end{center}
\vspace{-0.5cm}
\caption{Schematic illustration of Rabi oscillations in a $S=\frac{1}{2}$ pair for $J=0$ (a), and for strong exchange (b); $T_{\s -}$, $T_{\s 0}$, $T_{\s +}$ are the triplet states of the pair.}
\label{fig:spinstates}
\end{figure}
In recent years, various experimental~\cite{BoehmeC60,BoehmeNature,Brandt,BoehmeMain,Wrong}  studies focusing on the nature of spin--dependent charge carrier transport and recombination  processes have been conducted by using a coherent magnetic resonant spin manipulation. The idea of these pulsed electrically and optically detected magnetic resonance experiments (pEDMR, pODMR, respectively) is to identify an induced coherent spin propagation by observables that are directly controlled by coherent spin motion. Typically, for pEDMR and pODMR, a powerful oscillating driving field
is applied in magnetic resonance to the explored spin system, and the resulting Rabi oscillation is then observed electrically or optically. From the Rabi oscillation components, information about the Hamiltonian of the propagating system and, therefore, about its physical nature can be obtained. While this experimental approach is similar to the way coherent spin motion is observed with conventional pulsed electron paramagnetic resonance (pEPR), the different observables prohibit a direct comparison of pEPR and pEDMR/pODMR experiments. This difference of observables has been subject of a number of theoretical studies\cite{Boehme0,Boehme1,Boehme2,Boehme3,Weakexchange} in recent years. Most experimental and theoretical studies have focused on spin--selection rules induced by the Pauli blockade. Pauli blockade exists when a transition between two localized paramagnetic, singly occupied electron states into one doubly occupied singlet states is controlled by the pair state of the two $S=1/2$ spins before the transition. This so called intermediate pair model was first described by Kaplan, Solomon and Mott\cite{Mott} in 1978. Based thereon, successful descriptions of spin--dependent processes observed with EDMR and ODMR and in particular for pEDMR and pODMR experiments have been possible. Most of these theoretical studies used numerical methods in order to scrutinize experimental insights. This approach however does not allow the derivation of analytical expressions needed for the fit of experimental data and which can also limit fundamental qualitative understanding. Only recently, first analytical descriptions of coherently controlled spin--dependent intermediate pair transitions rates have been derived in Ref.~\onlinecite{Weakexchange}, this work however applies only to intermediate pairs with negligible exchange and spin--dipolar interaction.

When intermediate spin $S=1/2$ pairs, consisting of two pair partners $a$ and $b$, are weakly coupled, then the Rabi oscillations in each partner
take place independently as illustrated in Fig.~\ref{fig:spinstates}a.
The spin--Rabi oscillation frequencies of the pair partners will then become
\begin{eqnarray}
\label{sA}
s_{\s a}=(\delta_{\s a}^2+\Omega_{\s R}^2)^{1/2},~~
s_b=(\delta_{\s b}^2+\Omega_{\s R}^2)^{1/2}\!,
\end{eqnarray}
where $\Omega_{\s R}=\gamma B_{\s 1}$ is the Rabi nutation frequency,
$B_{\s 1}$ is the magnitude of the driving ac magnetic field, and $\gamma$ is gyromagnetic ratio;  $\delta_{\s a}=\omega_{\s a}-\omega$, and $\delta_{\s b}=\omega_{\s b}-\omega$ are the differences (the so called detuning) of the pair partners Larmor frequencies $\omega_{\s a}$ and $\omega_{\s b}$ from the excitation frequency $\omega$ (see Fig.~\ref{fig:spinstates}a).
For independent oscillations, the Rabi spectrum, ${\bf\mathrm{F}}(s)$, of electrically detected magnetic resonance will contain the lines
\begin{eqnarray}
\label{frequencies}
s_{\s 0}=|s_{\s a}-s_{\s b}|, ~~~ s_{\s 1}^{\s a}=s_{\s a}, ~~~s_{\s 1}^{\s b}=s_{\s b},~~~
s_{\s 2}=s_{\s a}+s_{\s b}.
\end{eqnarray}
The lines  $s_{\s 1}^{\s a}$ and $s_{\s 1}^{\s b}$ correspond to
precession of one of the two pair partners, whereas the $s_{\s 2}$-line\cite{Boehme0,BoehmeMain,Boehme1}
(and also the $s_{\s 0}$-line\cite{Weakexchange}  ) originates from coherent precession of {\em both} pair partners around $B_{\s 1}$ in the rotating frame, they are beat Rabi--beat components due to the relative spin motion of the two pair partners.

In the present paper we address the question of how the process in Fig.~\ref{fig:spinstates}a gets modified when the pair partners are coupled by strong exchange $J\gg \Omega_{\s R}, \delta_{\s a}, \delta_{\s b}$ and non--negligible spin--dipolar interaction. We demonstrate that, in the limit of strong exchange, the Rabi oscillations  are intrinsically collective and proceed according to the following scheme
\begin{equation}
\label{scheme}
\downarrow \downarrow \Leftrightarrow \frac{1}{\sqrt{2}}(\downarrow \uparrow+\uparrow\downarrow)
\Leftrightarrow \uparrow\uparrow.
\end{equation}
Fig.~\ref{fig:spinstates}b shows an illustration of this scheme which does not involve the singlet state,
$\frac{1}{\sqrt{2}}(\downarrow \uparrow-\uparrow\downarrow)$, which, as we demonstrate below, gets decoupled in the domain
$J\gg \Omega_{\s R}, \delta_{\s a}, \delta_{\s b}$.

We will show that, in the limit of large $J$, the spectrum of the Rabi oscillations is governed by the interplay of the Rabi frequency,
$\Omega_{\s R}$,  dipole-dipole interaction magnitude, $D$, and
the average detuning, $\delta$. As anticipated for a two $S=1/2$ system which gradually turns into one $s=1$ system with increasing $J$, the magnitude, $J$, of the exchange coupling drops out from the theory and the spectrum is governed by a {\em single} dimensionless  combination of parameters, $\Omega_{\s R}$, $D$, and $\delta$.
From previous numerical studies of exchange coupled intermediate pairs~\cite{Boehme2} it is known that due to the change of the two $S=1/2$ system into one $S=1$ system, EDMR and ODMR induced rate changes becomes negligibly small. Qualitatively, this can be understood by the realization that in presence of large $J$, magnetic resonance will always change triplet states into triplet states and the singlet to triplet ratio is therefore not changed. However, the results presented here can still be of significance as long as $J$ is large but not many orders of magnitude larger than dipolar or the Rabi nutation frequency.

Our main finding is that, upon the change of this parameter, the spectrum, ${\bf\mathrm{F}}(s)$, exhibits a non-trivial evolution. Peculiarity of ${\bf\mathrm{F}}(s)$ manifests itself in the behavior of the Rabi spectral lines at small $\Omega_{\s R}\lesssim D$.
For  two-level systems, the frequencies of oscillations always {\em grow} with increasing $\Omega_{\s R}$.
We find that, for the dipole-dipole coupled system in Fig.~\ref{fig:spinstates}b, the Rabi spectrum contains three
frequencies one of which {\em decreases} with $\Omega_{\s R}$.
Upon subsequent increase of  $\Omega_{\s R}$, this frequency passes through a minimum and grows as $s\approx \Omega_{\s R}$
at large $\Omega_{\s R}\gg D$. In addition, we find that the behavior
of this spectral line with detuning,  $\delta$, also exhibits
a minimum. Moreover, we find that a minimum in the position of the Rabi spectral line is accompanied by a {\em maximum} in its intensity. The definition of the line intensity pertinent to electrically detected  magnetic-resonance experiments
\cite{BoehmeC60,BoehmeNature,Brandt,BoehmeMain,Wrong,BoehmeSilicon1,BoehmeSilicon2,
Brandt2,BoehmeDifferentiation}
is given below. Finally, we demonstrate that the disorder with r.m.s., $\Delta$, in Larmor frequencies of the pair partners leads to a Gaussian
broadening of the Rabi spectral lines, and express the corresponding widths
in terms of $\Delta$, $\Omega_{\s R}$, $D$, and $\delta$.

In the previous numerical studies of the of the Rabi oscillation Fourier spectra\cite{Boehme1,Boehme2,Boehme3}, the frequencies of oscillations were found for various sets of parameters and different relations between $J$, $\Omega_{\s R}$, $\delta_{\s a}$ and $\delta_{\s b}$. Here, we restrict our consideration to the domain of large $J$, but within this domain our treatment is fully analytical.

The paper is organized as follows: In Sect. II we relate the time-dependent populations of different spin states involved in the Rabi oscillations of pairs to the observable quantity, namely,  photoconductivity measured during pEDMR experiments. In Sect. III we analyze the  quasienergies of a resonantly driven spin pair in the limit of strong exchange. In Sect. IV general expressions for positions and intensities of the Rabi spectral lines are derived. These expressions are analyzed in Sect. IV. In Sect. V we discuss the relation of spin-Rabi oscillations in a coupled pair to the excitonic Rabi oscillations in quantum dot molecules.

\section{pulsed EDMR techniques and the intensities of the Rabi spectral lines}
For pEDMR experiments, a samples' conductivity change, $\Delta \sigma$, is measured upon application of a short resonant magnetic resonant pulse~\cite{Boehme0}. More specifically, the dynamics, $\Delta \sigma(t)$, of the return of conductivity to the steady state after the pulse ends is
measured as a function of the pulse duration, $\tau$. This duration is much shorter than all intrinsic times, so that the change of conductivity during the interval, $\tau$,
is negligible. Dependence of $\Delta \sigma(t)$ on $\tau$ originates from the fact that the pulse rotates the spins of the pair partners. On the other hand, the spin state of the pair serves as initial condition for the process of the conductivity recovery. In this way, PEDMR measurements provide information about the Rabi oscillations within the pair of spins.


A very important observation made in  Ref. \onlinecite{Boehme0} is that the contribution to photoconductivity, $\Delta\sigma$, comes from  specific spin configurations both in
initial and in final states. More specifically, when thermal polarization is negligibly small and the system is in a steady state, the initial state of the pair at the moment,  $t=0$, of application of the microwave pulse is either $\downarrow\downarrow$, or $\uparrow\uparrow$, with equal probability. If the system was initially in $\downarrow\downarrow$, the contribution to $\Delta\sigma$ is proportional to $|A^{\s \downarrow\downarrow}_{\s \downarrow\downarrow}(\tau)|^2+ |A^{\s \downarrow\downarrow}_{\s \uparrow\uparrow}(\tau)|^2$, where $A_{\s \downarrow\downarrow}^{\s \downarrow\downarrow}(\tau)$ and $A_{\s \uparrow\uparrow}^{\s \downarrow\downarrow}(\tau)$ are the amplitudes to find the system, respectively, in
 $\downarrow\downarrow$ and $\uparrow\uparrow$ at time $\tau$, which is the
duration of the pulse. The upper indices indicate that the amplitudes are calculated with initial
 condition that at $t=0$ the system is in $\downarrow\downarrow$.
Correspondingly, if at $t=0$ the system was in $\uparrow\uparrow$, the contribution to $\Delta\sigma$ is proportional to $|A^{\s \uparrow\uparrow}_{\s \downarrow\downarrow}(\tau)|^2+ |A^{\s \uparrow\uparrow}_{\s \uparrow\uparrow}(\tau)|^2$. Therefore, the quantity, $\Delta\sigma$, measured by PEDMR techniques should be identified with the following combination of quantum-mechanical probabilities
\begin{equation}
\label{sum}
\Delta\sigma(\tau) \propto \Bigl[|A^{\s \downarrow\downarrow}_{\s \downarrow\downarrow}(\tau)|^2+ |A^{\s \downarrow\downarrow}_{\s \uparrow\uparrow}(\tau)|^2+|A^{\s \uparrow\uparrow}_{\s \downarrow\downarrow}(\tau)|^2+ |A^{\s \uparrow\uparrow}_{\s \uparrow\uparrow}(\tau)|^2\Bigr].
\end{equation}
Each of the probabilities contains three oscillating components of the form $\cos s\tau$. Thus, the Fourier analysis of measured $\Delta\sigma(\tau)$ should reveal three peaks. We define the intensity of the Rabi spectral line as a magnitude of the corresponding oscillating component in the sum Eq.~(\ref{sum}).


\section{Quasienergies of the driven system}
We start from the  Hamiltonian of the pair
\begin{align}
&\label{H}
{\hat {H}}=
\omega_{\s a}S_{\s a}^{\s z}+\omega_{\s b}S_{\s b}^{\s z}+
2\Omega_{\s R}(S_{\s a}^{\s x}+S_{\s b}^{\s x})\cos\omega t
&
\nonumber \\
&\hspace{0.8cm}-J\hat{{\bf S}}_{{\s a}}\cdot\hat{{\bf S}}_{{\s b}}
-D\big(3S_{\s a}^{\s z}S_{\s b}^{\s z}-\hat{\bf S}_{\s a}\cdot\hat{\bf S}_{\s b}\big),&
\end{align}
where the first three terms represent the Hamiltonian of the ac driven
pair partners and the last two terms describe the intra-pair exchange
and dipole-dipole interactions.

To study the time evolution of the spin-pair one has to solve the Schr{\"o}dinger equation, $ i\frac{\partial}{\partial t}\Psi(t)= \hat{H}\Psi(t)$, for the four-component wave function
\begin{equation}
\label{psi}
\Psi(t)=\big\{A_{\uu}(t),\,A_{\dd}(t),\,A_{\du}(t),\,A_{\ud}(t)\big\}
\end{equation}
of the amplitudes of different spins states. These amplitudes satisfy the
following system of equations
\begin{align}
\label{S.E.}
&i\frac{\partial A_{\dd}}{\partial t}=-
\frac{1}{2}\left(J+D+\omega_{\s a}+\omega_{\s b}\right)A_{\dd}
&
\nonumber\\
&\hspace{1.3cm}
+\Omega_{\s R}\cos \omega t
(A_{\ud}+A_{\du}),&
\nonumber \\
&i\frac{\partial A_{\uu}}{\partial t}=-
\frac{1}{2}\left(J+D-\omega_{\s a}-\omega_{\s b}\right)A_{\uu}&
\nonumber \\
&\hspace{1.3cm}+\Omega_{\s R}\cos \omega t
(A_{\ud}+A_{\du}),&
\nonumber \\
&i\frac{\partial A_{\du}}{\partial t}=
\left(\frac{D}{2}-\delta_{\s 0}\right)A_{\du}-\frac{1}{2}(J-D)A_{\ud}&
\nonumber \\
&\hspace{1.3cm}+\Omega_{\s R}\cos \omega t
(A_{\dd}+A_{\uu}),&
\nonumber \\
&i\frac{\partial A_{\ud}}{\partial t}=
\,\,\left(\frac{D}{2}+\delta_{\s 0}\right)A_{\ud}-\frac{1}{2}(J-D)A_{\du}&
\nonumber \\
&\hspace{1.3cm}+\Omega_{\s R}\cos \omega t
(A_{\dd}+A_{\uu}),&
\end{align}
where the asymmetry parameter, $\delta_{\s 0}$ is defined as
\begin{equation}
\label{delta0}
\delta_{\s 0}=\frac{\delta_{\s a}-\delta_{\s b}}{2}=\frac{\omega_{\s a}-\omega_{\s b}}{2}.
\end{equation}
The quasienergies, $\chi$, of the system of equations Eq. \eqref{S.E.} are introduced in a standard way, by using the following substitutions
\begin{align}
\label{subst}
&A_{\dd}=a_{\dd}e^{-i(\chi-\omega) t},
\hspace{1.3cm}
A_{\uu}=a_{\uu}e^{-i(\chi+\omega) t},&\nonumber \\
&A_{\ud}=a_{\ud}e^{-i\chi t},
\hspace{1.85cm}
A_{\du}=a_{\du}e^{-i\chi t},&
\end{align}
and employing the rotating-wave approximation.
Then the system Eq. (\ref{S.E.}) reduces to the following
system of algebraic equations
\begin{align}
\label{4eqns}
&\hspace{0.5cm}\left ( \chi-\delta+\frac{J+D}{2} \right)a_{\dd}=
\frac{\Omega_{\s R}}{2}\left(a_{\ud}+a_{\du}\right),&
\nonumber \\
&\hspace{0.5cm}\left ( \chi+\delta+\frac{J+D}{2} \right)a_{\uu}=
\frac{\Omega_{\s R}}{2}\left(a_{\ud}+a_{\du}\right),&
\nonumber \\
&\left ( \chi+\delta_{\s 0} -\frac{D}{2}\right)a_{\ud}+\frac{1}{2}(J-D)a_{\du}=
\frac{\Omega_{\s R}}{2}\left(a_{\uu}+a_{\dd}\right),&
\nonumber \\
&\left ( \chi-\delta_{\s 0}-\frac{D}{2}  \right)a_{\du}+\frac{1}{2}(J-D)a_{\ud}=
\frac{\Omega_{\s R}}{2}\left(a_{\uu}+a_{\dd}\right),&
\end{align}
where we introduced the detuning parameter
\begin{equation}
\label{detuning}
\delta=\frac{\delta_{\s a}+\delta_{\s b}}{2}=\frac{\omega_{\s a}+\omega_{\s b}}{2}-\omega.
\end{equation}
The system Eq. \eqref{4eqns} can be reduced to two coupled equations for the amplitudes $a_{\ud}$ and $a_{\du}$, which read
\begin{flalign}
\label{coupled1}
&\left[\tilde{\chi}-D+\delta_{\s 0}-\frac{J}{2}-
\frac{\Omega_{\s R}^2}{2} \left(\frac{\tilde{\chi}}{\tilde{\chi}^2-\delta^2}\right)
\right]a_{\ud}&
\nonumber \\
&\hspace{2.5cm}=\frac{1}{2}\left[D-J+\Omega_{\s R}^2\left(\frac{\tilde{\chi}}{\tilde{\chi}^2-\delta^2}\right)
 \right]a_{\du},&
\end{flalign}
\begin{flalign}
\label{coupled2}
&\left[\tilde{\chi}-D-\delta_{\s 0}-\frac{J}{2}-
\frac{\Omega_{\s R}^2}{2} \left(\frac{\tilde{\chi}}{\tilde{\chi}^2-\delta^2}\right)
\right]a_{\du}&
\nonumber \\
&\hspace{2.5cm}=\frac{1}{2}\left[D-J+\Omega_{\s R}^2\left(\frac{\tilde{\chi}}{\tilde{\chi}^2-\delta^2}\right)
 \right]a_{\ud},&
\end{flalign}
where $\tilde{\chi}=\chi+\frac{1}{2}(J+D)$.
Multiplying
Eqs. (\ref{coupled1}) and (\ref{coupled2}) gives the following quartic equation for the quasienergies
\begin{equation}
\label{quartic}
\left[\tilde{\chi}-\frac{3}{2}D-\left(\frac{\delta_{\s 0}^2}{\tilde{\chi}-J-\frac{D}{2}}\right)\right]
\big(\tilde{\chi}^2-\delta^2\big)=\Omega_{\s R}^2\tilde{\chi}.
\end{equation}
The form Eq. (\ref{quartic}) of the characteristic equation  explains on the quantitative level the
statement made in the Introduction that, at large $J$, the magnitude
of exchange drops out from the theory. Indeed, if the asymmetry
parameter $\delta_{\s 0}$ is much smaller than $(JD)^{1/2}$, $(J \Omega_{\s R})^{1/2}$, the last terms in the brackets,
containing $J$ in the denominator, can be neglected. More precisely,
under the condition $\delta_{\s 0}\ll (JD)^{1/2}, (J \Omega_{\s R})^{1/2}$, the quasienergy  $\tilde{\chi}\approx J+\frac{D}{2}$, corresponding to the singlet, is much bigger than three other quasienergies. As a result,  only three triplet states $A_{\uu}$, $A_{\dd}$, and $\frac{1}{\sqrt{2}}(A_{\ud}+A_{\du})$ participate in the Rabi oscillations. On the qualitative level, decoupling of the
singlet state, $\frac{1}{\sqrt{2}}(A_{\ud}-A_{\du})$, originates from
the fact that exchange does not mix this state with components of
the triplet.

In the limit of strong exchange Eq. \eqref{quartic}  reduces to
a cubic equation
\begin{equation}
\label{cubic}
\big(\tilde{\chi}^2-\delta^2\big)\left(\tilde{\chi}-\frac{3}{2}D\right)
=\Omega_{\s R}^2\tilde{\chi}.
\end{equation}
Obviously, for $D\ll \Omega_{\s R}$, $\delta$,
we recover the conventional Rabi oscillations
\begin{equation}
\chi_{\s 1} =0,~~ \chi_{\s 2,3} =\pm \sqrt{\delta^2+\Omega_{\s R}^2}.
\end{equation}
In the opposite limit, when $D\gg \Omega_{\s R},\, \delta$, we have
\begin{equation}
\label{chi23}
\chi_{\s 1} =\frac{3}{2}D,~~~
\chi_{\s 2,3} =\frac{\Omega_{\s R}^2}{3D}\pm \sqrt{\delta^2+\left(\frac{\Omega_{\s R}^2}{3D}\right)^2},
\end{equation}
so the difference $\chi_{\s 2}-\chi_{\s 3}$, which is the position of the lowest spectral
 line, behaves as $s=\frac{2}{3D}\Omega_{\s R}^2$. The positions of other two lines are close to $s=\frac{3}{2}D$.

To assess the intermediate regime, $\delta \sim D$,   we rewrite the cubic equation, Eq. (\ref{cubic}), by using the dimensionless variable,
\begin{equation}
\label{subchi}
\tilde{\chi}= \upsilon\eta +\frac{D}{2},
\end{equation}
where
\begin{equation}
\label{upsilon}
\upsilon=\sqrt{\frac{3}{4}D^2+\Omega_{\s R}^2 +\delta^2}.
\end{equation}
Then it assumes the form
\begin{equation}
\label{eq:eta}
\eta^3-\eta + f=0.
\end{equation}
We see that quasienergies, ${\tilde \chi}_{\s i}$, are determined by a {\em single} dimensionless
combination of parameters $\Omega_{\s R}$, $D$, and $\delta$,
defined as
\begin{equation}
\label{f}
f=-\frac{\frac{1}{2}D^2-2\delta^2+\Omega_{\s R}^2}{2\left(\frac{3}{4}D^2+\delta^2+\Omega_{\s R}^2\right)^{3/2}}D.
\end{equation}
\begin{figure}[t]
\begin{center}
\includegraphics[width=75mm]{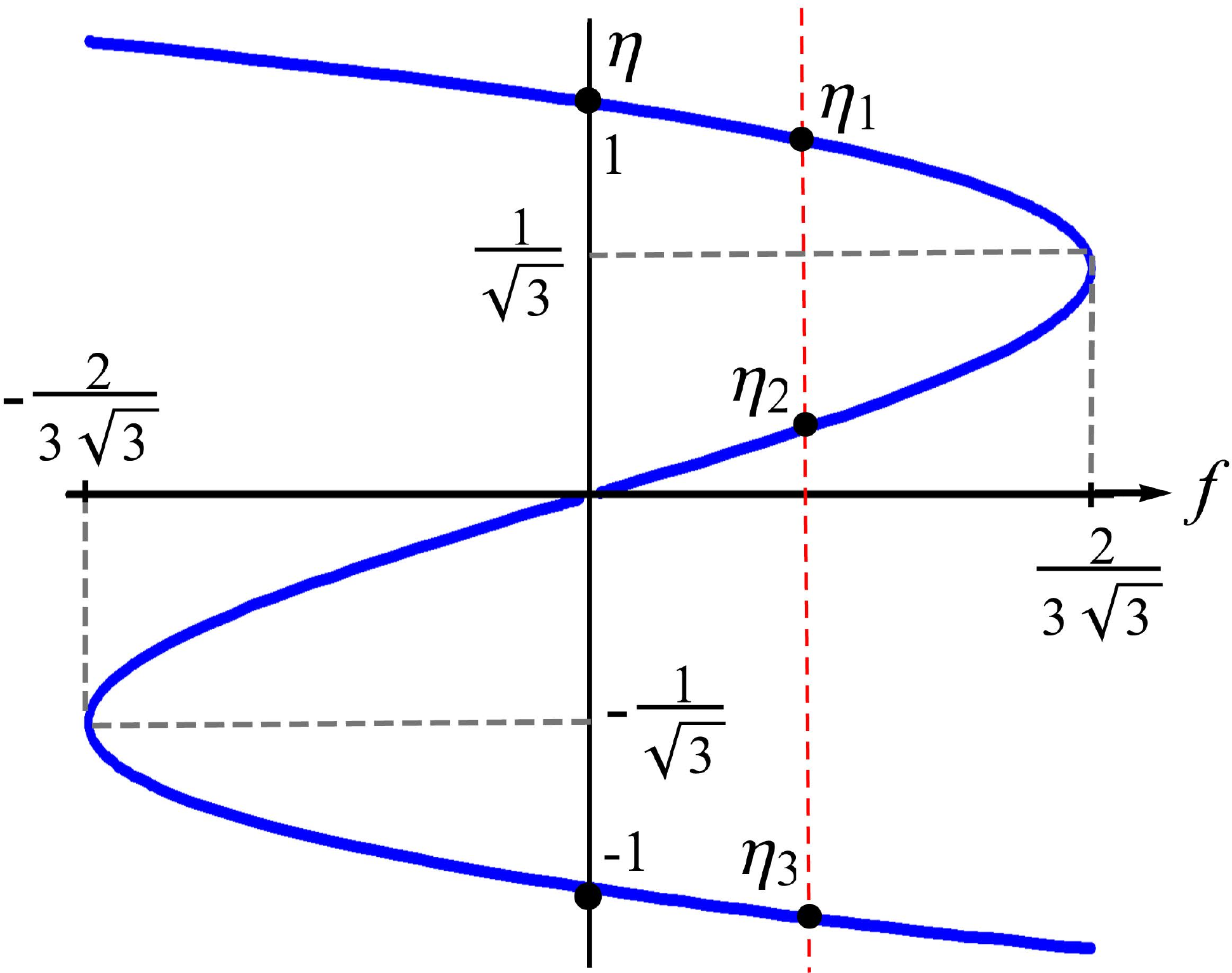}
\end{center}
\vspace{-0.5cm}
\caption{Graphic solution of the cubic equation Eq. (\ref{eq:eta}); $\eta_1$, $\eta_2$, $\eta_3$ are the roots of a given $f$.}
\label{fig:cubic}
\end{figure}
\begin{figure}[t]
\begin{center}
\includegraphics[width=75mm]{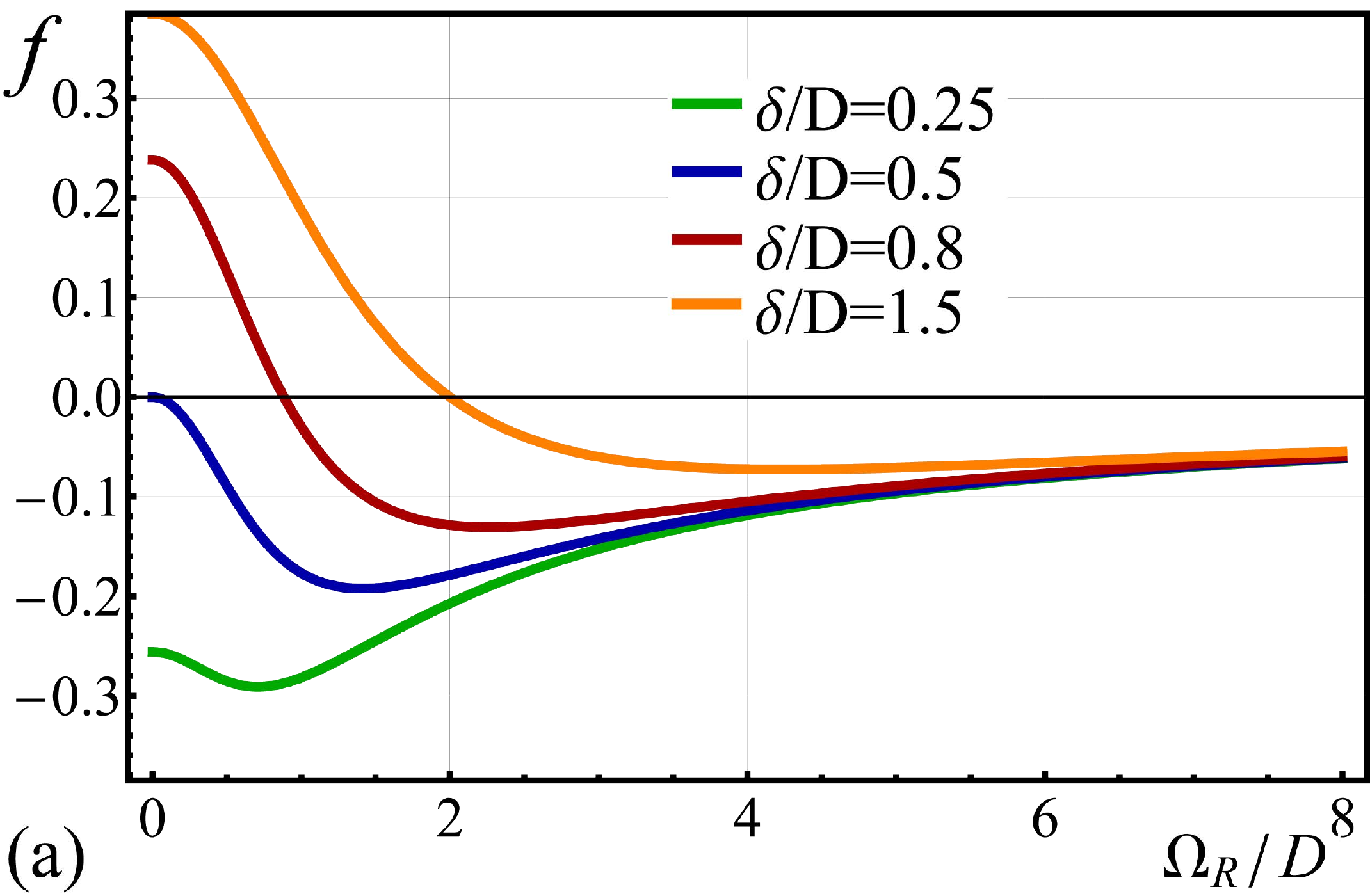}
\includegraphics[width=75mm]{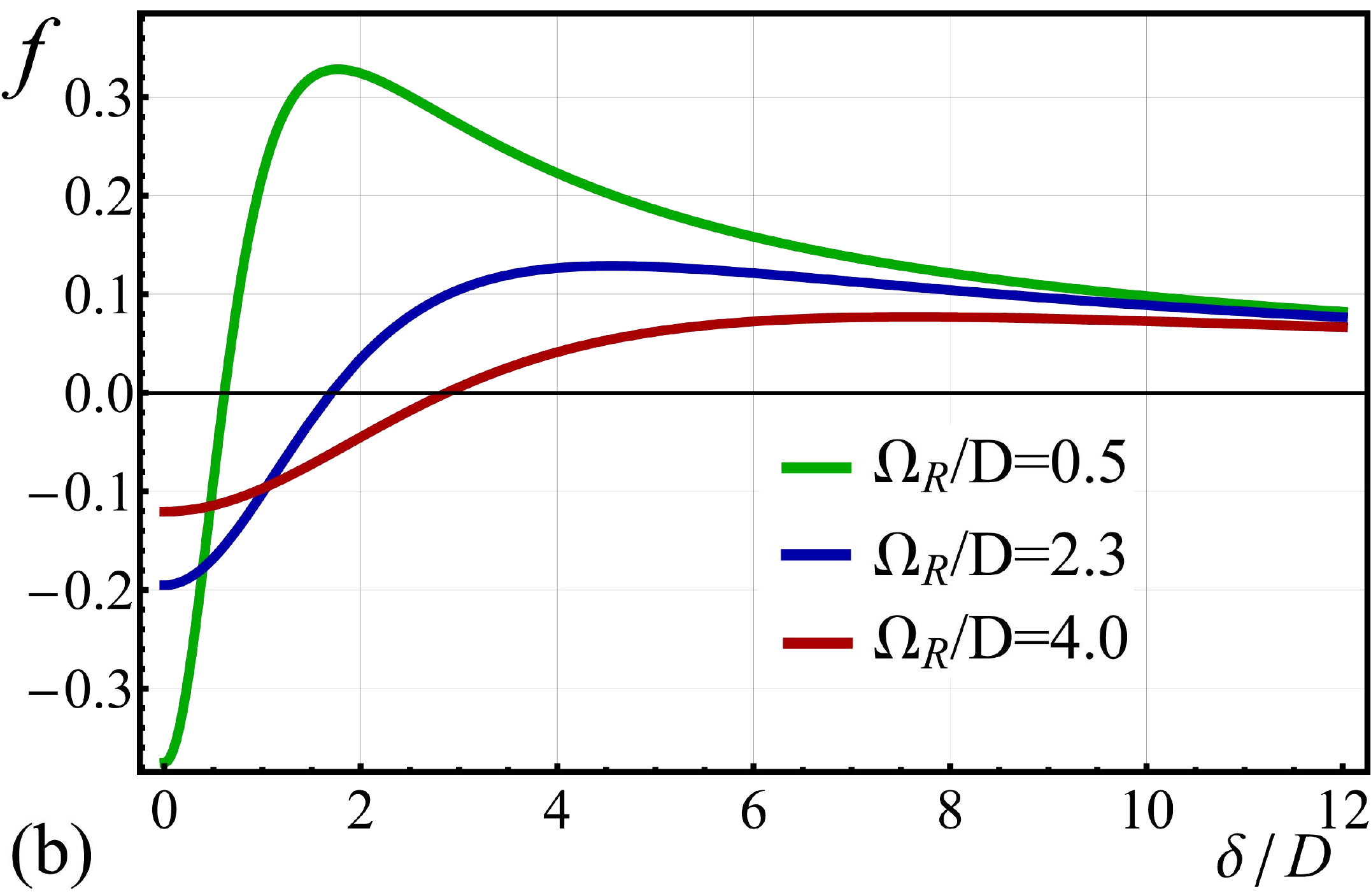}
\end{center}
\vspace{-0.3cm}
\caption{Dimensionless parameter $f$ is plotted from Eq. \eqref{f} as a function of dimensionless Rabi frequency $\Omega_{\s R}/D$ (a) and dimensionless detuning $\delta/D$ (b). In (a) the maximum moves to the right with increasing $\delta/D$. Whereas, in (b) the minimum moves to the right with increasing $\Omega_{\s R}/D$. }
\label{fig:f}
\end{figure}
It is easy to check that for any $D,\,\delta,$ and $\Omega_{\s R}$ the parameter $f$ resides
within the interval $- \frac{2}{3\sqrt{3}}, \frac{2}{3\sqrt{3}}$, which ensures that all
three roots, $\eta_{\s i}$, are real.
Graphic solution of Eq. \eqref{eq:eta} is illustrated in Fig.~\ref{fig:cubic}.
Analytic expressions for the roots, $\eta_{\s i}$, are the following
\begin{align}
\label{eigenvalues}
&&\eta_1= -\mbox{sgn}(f)\frac{2}{\sqrt{3}}\cos\left(\frac{\psi}{3}\right)\!,\hspace{0.9cm}
\nonumber \\
&&\eta_2= -\mbox{sgn}(f)\frac{2}{\sqrt{3}}\cos\left(\frac{\psi}{3}+\frac{2\pi}{3}\right)\!,
\nonumber \\
&&
\eta_3=-\mbox{sgn}(f)\frac{2}{\sqrt{3}}\cos\left(\frac{\psi}{3}-\frac{2\pi}{3}\right)\!,
\end{align}
where the phase $\psi$ is determined as
\begin{equation}
\label{psi}
\psi=\arctan\left( \frac{1}{f}\sqrt{\frac{4}{27}-f^2}\right)\!.
\end{equation}
To find the quasienergies, ${\tilde\chi}_{\s i}$ for given values of $D$, $\delta$, and $\Omega_{\s R}$, one has to calculate parameter $f$ from Eq. \eqref{f}, substitute it into Eq. \eqref{eigenvalues} for $\eta_i$, and, finally, substitute $\eta_i$ into Eq. \eqref{subchi}.
The evolution of ${\tilde\chi}_{\s i}$ with $\Omega_{\s R}$ and $\delta$ is governed by the
dependence of $f$ on these parameters, which can be tuned externally. Fig.~\ref{fig:f} illustrates
that this evolution is quite nontrivial, namely, $f$ exhibits extrema both as function of
$\Omega_{\s R}$ and as a function of $\delta$.

\section{Positions and intensities of spectral lines}

The eigenvectors of the system Eq. (\ref{4eqns}) corresponding to the roots can be conveniently
cast in the form
\begin{equation}
\label{eigenvectors2}
X_{i}=	
	\begin{pmatrix}
	\vspace{0.2cm}
	\frac{ \Omega_{\s R} }{\sqrt{2}\big(\upsilon \eta_{\s i}+\frac{1}{2}D-\delta\big)}
	\\
	\vspace{0.2cm}
	 1 \\	
	\vspace{0.2cm}
	\frac{ \Omega_{\s R} }{\sqrt{2}\big(\upsilon \eta_{\s i}+\frac{1}{2}D+\delta\big)}	
	\end{pmatrix}\!\!.
\end{equation}
We are now in position to calculate the population of the state $\frac{1}{\sqrt{2}}(A_{\ud}+A_{\du})$ as a function of time.  The general expression for $|A_{\ud}+A_{\du}|^2$ can be written as
\begin{align}
\label{c squared}
&|A_{\ud}+A_{\du}|^2=C_1^2+C_2^2+C_3^2+
2C_1 C_2\, \cos \big[\upsilon(\eta_1-\eta_2) t\big]&
\nonumber \\
&+2C_1 C_3\, \cos\big[ \upsilon(\eta_1-\eta_3) t\big]
+2C_2 C_3\, \cos\big[ \upsilon(\eta_2-\eta_3) t\big]\!,&
\end{align}
 where the three constants $C_1$, $C_2$, and $C_3$, are determined from three initial conditions.
Assuming that at $t=0$ the
pair of spins is in $\downarrow\downarrow$ state, so that $A_{\dd}(0)=1$, $A_{\uu}(0)=0$, $(A_{\ud}(0)+A_{\du}(0))=0$, and solving the system of three linear
equations we find
\begin{align}
\label{IVconstants}
&C_1=\frac{\Omega_{\s R}\big(\upsilon\eta_1-\delta+\frac{D}{2}\big)}
{\sqrt{2} \upsilon^2(1-3\eta_1^2)},
~~~C_2=\frac{\Omega_{\s R}\big(\upsilon\eta_2-\delta+\frac{D}{2}\big)}
{\sqrt{2} \upsilon^2(1-3\eta_2^2)},&
\nonumber \\
&\hspace{2.5cm}C_3=\frac{\Omega_{\s R}\big(\upsilon\eta_3-\delta+\frac{D}{2}\big)}
{\sqrt{2} \upsilon^2(1-3\eta_3^2)}.&
\end{align}
\begin{figure}[h!]
\begin{center}
\includegraphics[width=78mm]{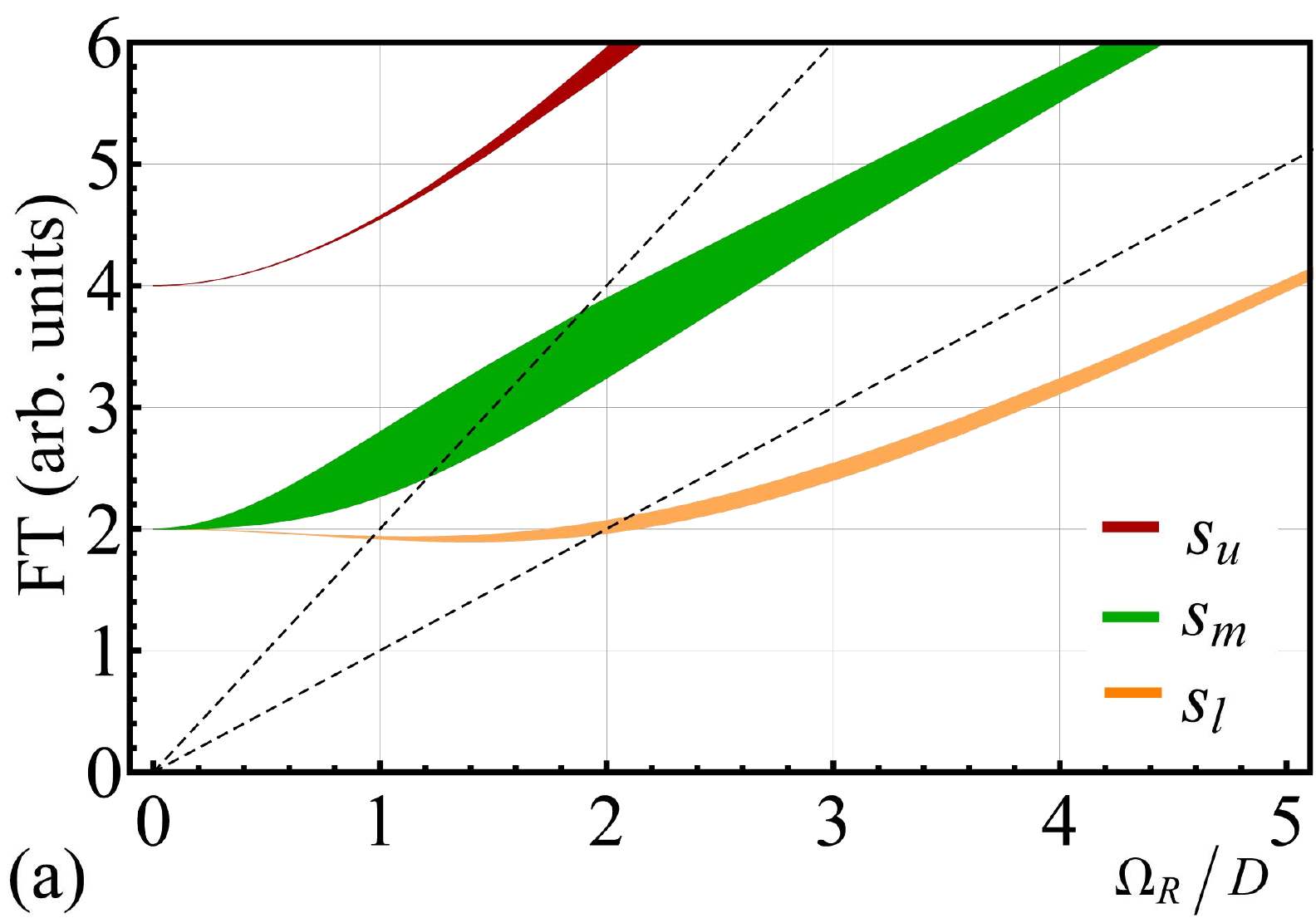}
\includegraphics[width=78mm]{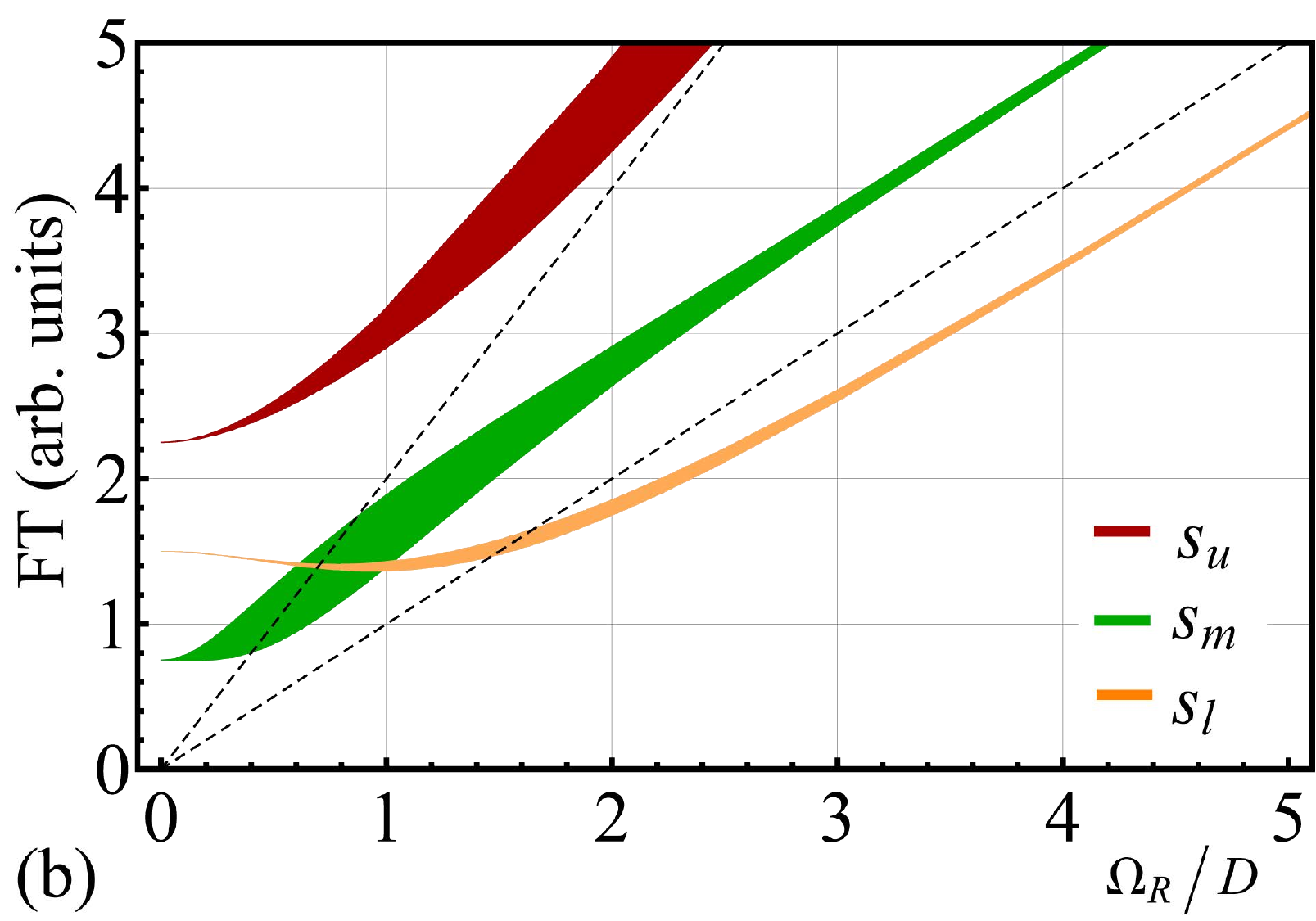}
\includegraphics[width=78mm]{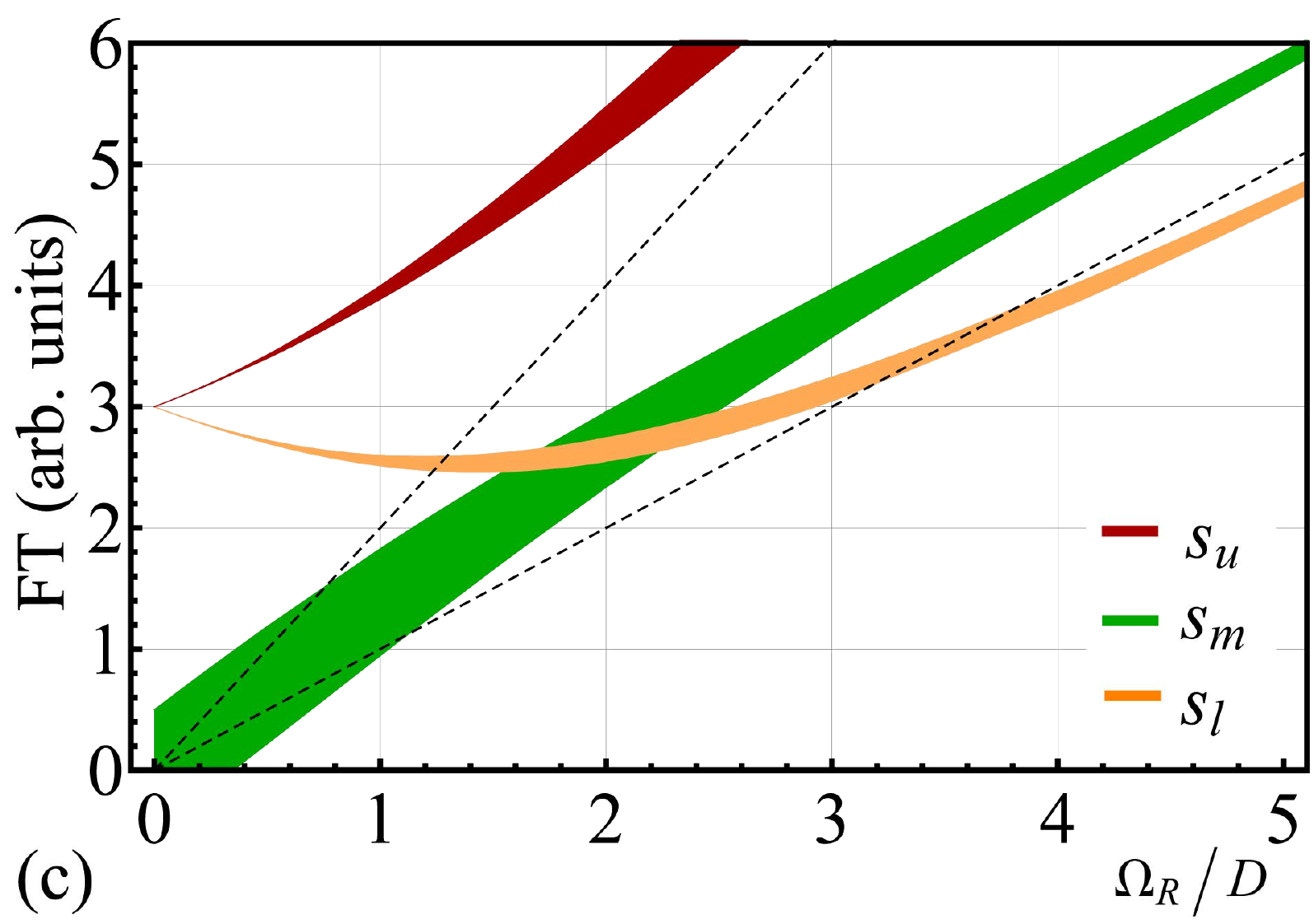}
\end{center}
\vspace{-0.7cm}
\caption{Spectra of the Rabi oscillations in the limit of strong exchange are plotted from Eq. \eqref{c squared} versus dimensionless Rabi frequency, $\Omega_{\s R}/D$, for three values of dimensionless detuning, $\delta/D=0.5$ (a), $\delta/D=0.75$ (b), and $\delta/D=1.5$ (c). The thickness of each line represents the corresponding peak intensity. Upon increasing $\Omega_{\s R}$ two lower peaks approach $s=\Omega_{\s R}$ (lower dashed line), while the upper peak approaches $s=2\Omega_{\s R}$ (upper dashed line).}
\label{fig:FT}
\end{figure}
At this point we note that the population Eq. (\ref{c squared}) is directly related to the photoconductivity, $\Delta\sigma$, Eq. \eqref{sum}. Indeed, $1-|A_{\ud}+A_{\du}|^2$ is  a probability to find the
system either in $\downarrow\downarrow$ state or in $\uparrow\uparrow$ state after
the time $t$.


The second contribution to $\Delta\sigma$ comes from realizations
in which the system is initially in the $\uparrow\uparrow$ state.
It is easy to see that this contribution can be obtained by simply changing $\delta$ by $-\delta$ in Eq. (\ref{IVconstants}). In terms of the Fourier transform, as it follows from Eqs. \eqref{sum} and \eqref{c squared},  $\Delta\sigma$ will contain the three peaks: upper, lower, and middle, with positions
\begin{eqnarray}
\label{FTfreq}
&s_{\s u}= \upsilon|\eta_1-\eta_2|,~~~
s_{\s m}= \upsilon|\eta_1-\eta_3|,&
\nonumber \\
&s_{\s l }= \upsilon|\eta_2-\eta_3|,&
\end{eqnarray}
and  corresponding intensities
\begin{eqnarray}
\label{intensities}
&{\bf\mathrm{F}}(s_{\s u})=2C_1(\delta) C_2(\delta)+2C_1(-\delta) C_2(-\delta),&
\nonumber\\
&{\bf\mathrm{F}}(s_{\s m})=2C_1(\delta) C_3(\delta)+2C_1(-\delta) C_3(-\delta),&
\nonumber\\
&{\bf\mathrm{F}}(s_{\s l})=2C_2(\delta) C_3(\delta)+2C_2(-\delta) C_3(-\delta).&
\end{eqnarray}
In the next Section we analyze how the peak positions and magnitudes evolve with
$\Omega_{\s R}$.
\section{Analysis and discussion}
In the previous section we derived analytical expressions for the
positions of the Rabi spectral lines and their intensities, see Eqs. \eqref{FTfreq} and
\eqref{intensities}. Below we analyze the evolution of the spectrum with increasing the
amplitude, $\Omega_{\s R}$, of the driving ac field and with detuning, $\delta$.

\subsection{Peak positions}
The positions of peaks in ${\bf\mathrm{F}}(s)$ as a function of $\Omega_{\s R}$ are plotted from Eqs. \eqref{upsilon}, \eqref{f}, and \eqref{eigenvalues}
in Fig.~\ref{fig:FT} for representative values of $\delta$.
The most interesting feature of ${\bf\mathrm{F}}(s)$-dependencies
is the behavior of the $s_{\s l}$-peak shown with yellow. At large $\Omega_{\s R}$
this peak is located below two other  peaks.
However, at small $\Omega_{\s R}$,  while the
$s_{\s u}$ and $s_{\s m}$ peaks grow monotonically
with $\Omega_{\s R}$, this peak either stays horizontal or even
{\em decreases} with $\Omega_{\s R}$.
Also the position of this peak at small $\Omega_{\s R}$ depends strongly on relation
between $\delta$ and $D$.
This can be understood from Eq. \eqref{cubic}.
At small $\Omega_{\s R}$ we have
$\tilde{\chi}_{\s 1}\approx \frac{3}{2}D$, and
$\tilde{\chi}_{\s 2,3}= \pm \delta$,
so that
\begin{equation}
\label{asymptotic}
s_{\s m}=\left|\frac{3}{2}D-\delta\right|,~~ s_{\s u}=\frac{3}{2}D+\delta,~~ s_l=2\delta.
\end{equation}
We see that for
$\delta \ll D$ or $\delta \gg D$
the peaks $s_{\s m}$ and $s_{\s u}$ are degenerate,
while for $\delta \approx \frac{3}{2}D$ the peaks $s_{\s u}$
and $s_{\s l}$ are degenerate.

The unusual behavior of $s_{\s l}$-peak with $\Omega_{\s R}$ can be understood from Figs.~\ref{fig:cubic} and \ref{fig:f}. First, as it follows from Eq.  \eqref{f},
parameter $f$ approaches zero  for large
$\Omega_{\s R}\gg \delta, D$. Then from Fig.~\ref{fig:cubic} we conclude that the roots of Eq. \eqref{eq:eta} approach $0$ and $\pm 1$ at large $\Omega_{\s R}$.
This translates into the following evolution of the peaks at large $\Omega_{\s R}$. Since in this regime  $\upsilon \approx \Omega_{\s R}$,
we find from Eq. \eqref{FTfreq}
\begin{equation}
s_{\s u}\approx 2\Omega_{\s R},~~s_{\s m}\approx s_{\s l}\approx \Omega_{\s R}.
\end{equation}
At this point we make the key observation
that parameter $f$ is
a {\em nonmonotonic} function of $\Omega_{\s R}$, as illustrated in Fig.~\ref{fig:f}a.
We see that all three curves have a maximum. This maximum shifts to the right with increasing $\delta$. Nonmonotonic dependence of $f$ on $\Omega_{\s R}$ affects the evolution of the roots of Eq. \eqref{eq:eta} with
$\Omega_{\s R}$. Indeed, to find these roots graphically,
one has to draw a vertical line in Fig.~\ref{fig:cubic}
corresponding to the abscissa equal to $f$ and find its
intersections with the blue curve. Since,
with increasing $\Omega_{\s R}$, the value of $f$
first increases and then decreases,
the dashed-red line in Fig.~\ref{fig:cubic} first
moves to the right and then
back to the left towards $f=0$.
This explains the nonmonotonic behavior
of the roots of Eq. \eqref{eq:eta}, and, subsequently, the non-monotonic behavior of the peak $s=s_{\s l}$ in Fig.~\ref{fig:FT}.
Note that, for large enough $\delta$, the value of
$f$ at small $\Omega_{\s R}$ is negative. Then,
upon increasing $\Omega_{\s R}$, the dashed line in Fig.~\ref{fig:cubic}
moves from $f<0$ to the left and crosses $f=0$.
Upon further increasing $\Omega_{\s R}$, parameter $f$
passes through a maximum at
\vspace{1cm}
\begin{figure}[h!]
\begin{center}
\includegraphics[width=77mm]{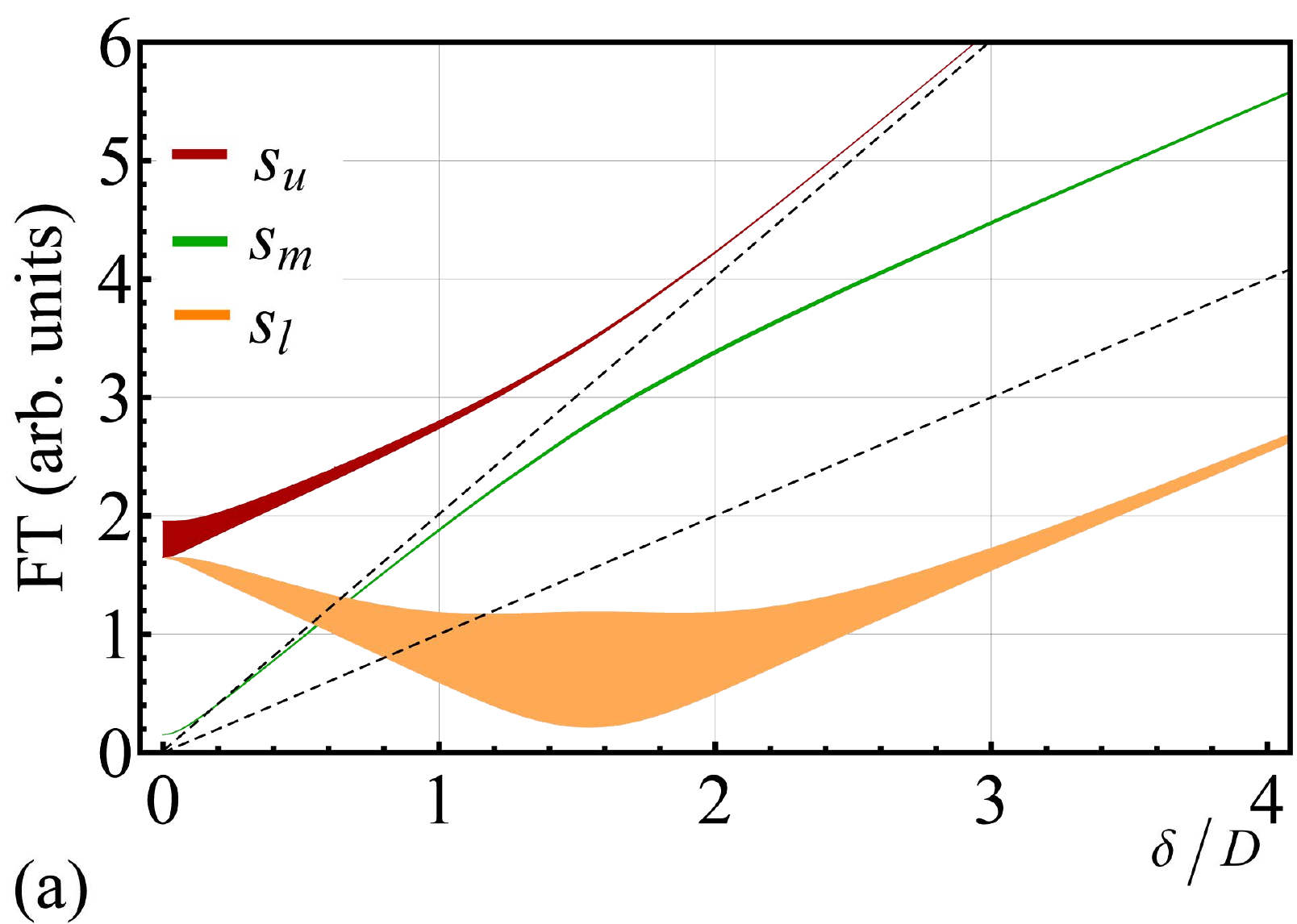}
\includegraphics[width=77mm]{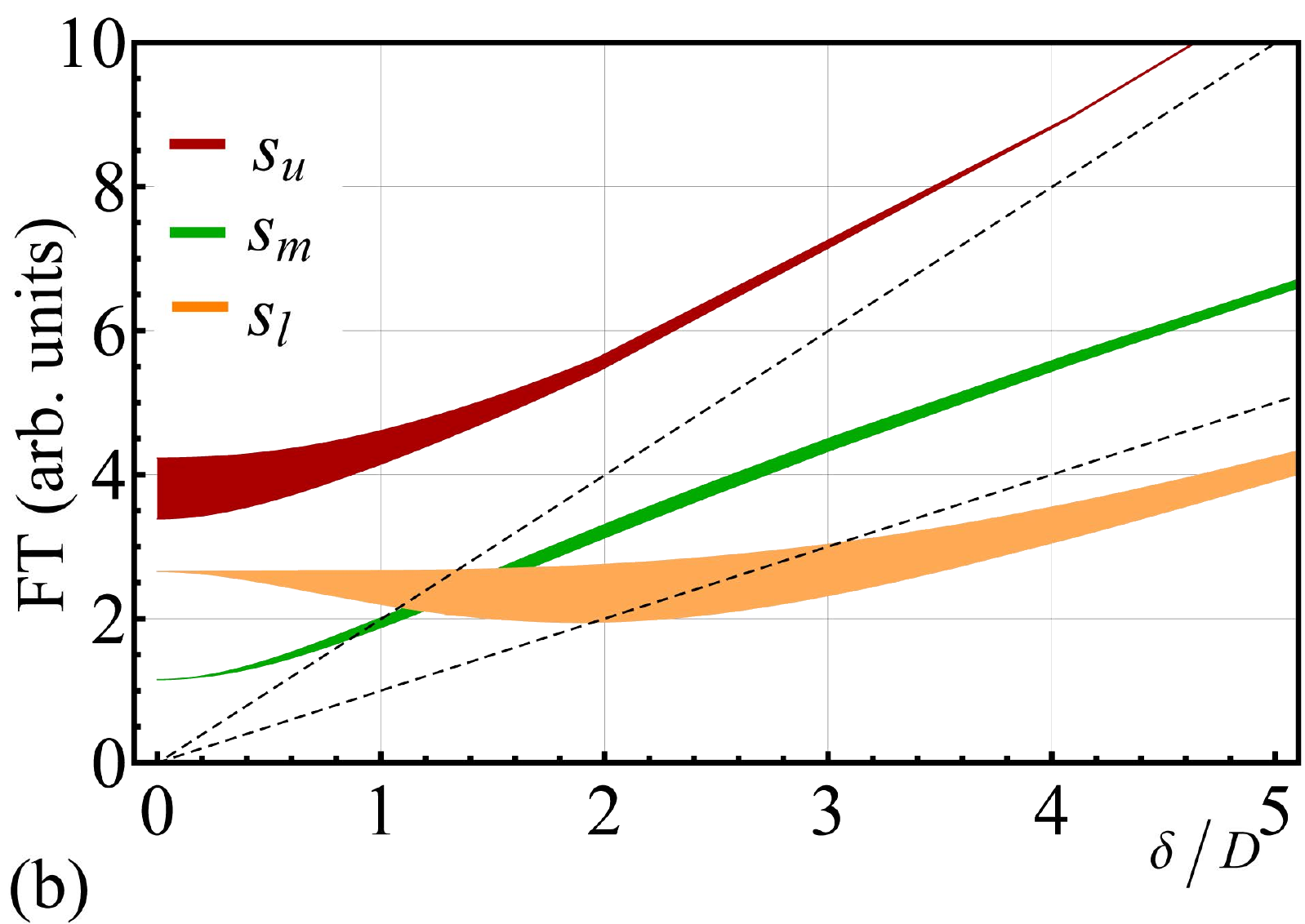}
\includegraphics[width=77mm]{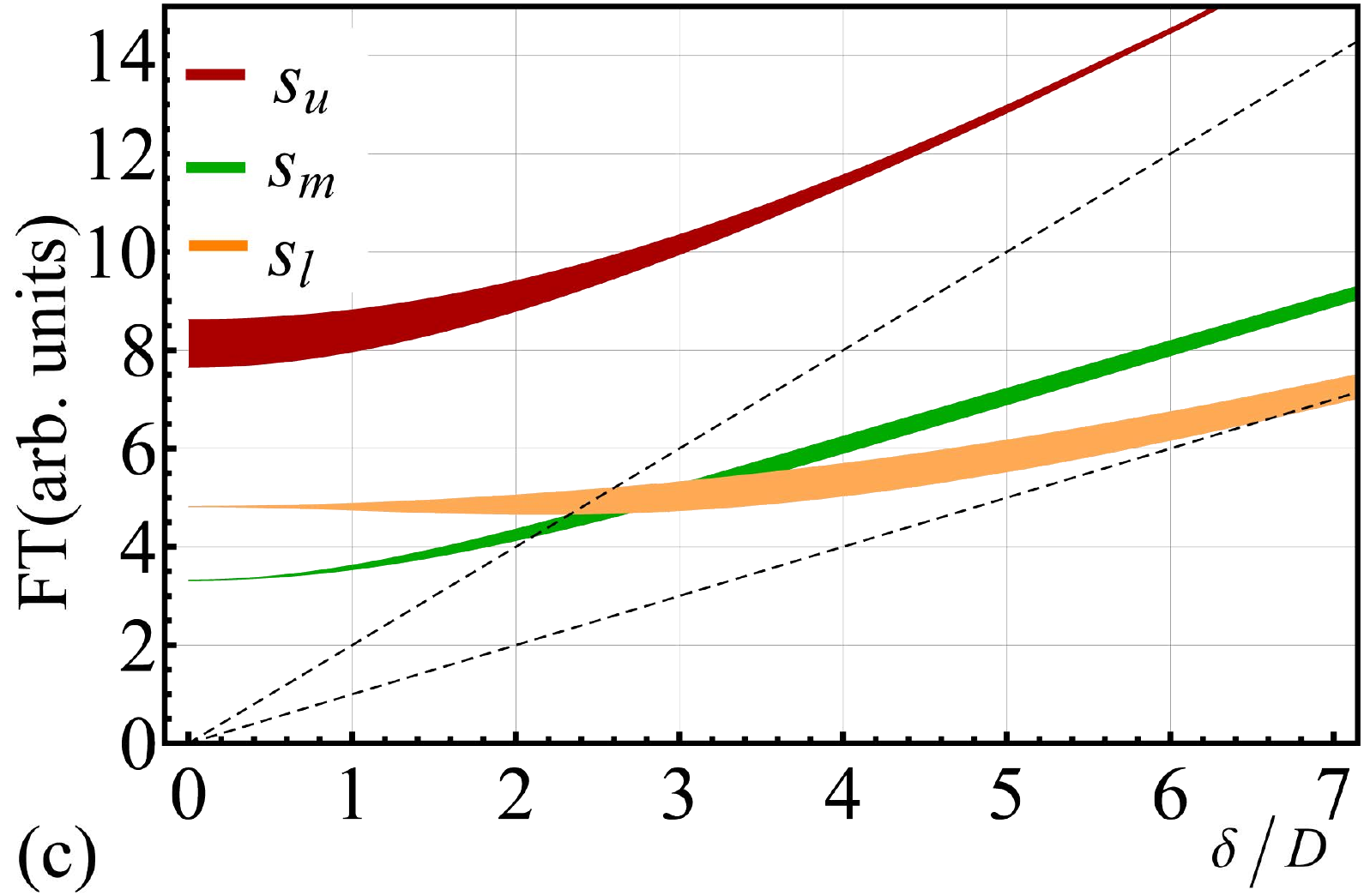}
\end{center}
\vspace{-0.7cm}
\caption{Spectra of the Rabi oscillations in the limit of strong exchange are plotted from Eq. \eqref{c squared} versus dimensionless detuning, $\delta/D$, for three values of dimensionless Rabi frequency, $\Omega_{\s R}/D=0.5$ (a), $\Omega_{\s R}/D=1.75$ (b), and $\Omega_{\s R}/D=4.0$ (c). Similar to Fig.~\ref{fig:FT}, the thickness of each line represents the corresponding peak intensity.
Upon increasing $\delta$ two lower peaks approach $s=\delta$ (lower dashed line), while the upper peak approaches $s=2\delta$ (upper dashed line).}
\label{fig:FTdelta}
\end{figure}
\begin{equation}
\Omega_{\s R}=2\sqrt{2}\delta,
\end{equation}
and returns
back to $f=0$.
This behavior corresponds to the most extended
interval in Fig.~\ref{fig:FT}b,c where $s_{\s l}(\Omega_{\s R})$ has a ``wrong"
slope.
Note that, for large enough $\delta$, the peaks $s=s_{\s l}(\Omega_{\s R})$ and $s=s_{\s m}(\Omega_{\s R})$ cross each other.
This crossing finds its natural explanation in the fact
that for large enough $\delta$ parameter $f$ turns to zero at
{\em finite} $\Omega_{\s R}$, namely at $\Omega_{\s R}=(2\delta^2-\frac{1}{2}D^2)^{1/2}$.

The evolution of the Rabi spectral lines with detuning, $\delta$, is illustrated
in Fig.~\ref{fig:FTdelta}. We see that, similarly to the $\Omega_{\s R}$-dependence,
the peak positions also evolve in  a nonmonotonic fashion. Such a behavior
can be readily accounted for by a nonmonotonic dependence of parameter $f$ on $\delta$. Indeed, as shown in Fig.~\ref{fig:f}b, the dependencies $f(\delta/D)$ pass through minimum at
\begin{equation}
\delta =\left(\frac{9}{4}D^2+\frac{7\Omega_{\s R}^2}{2}\right)^{1/2}
\end{equation}
for all values of $\Omega_{\s R}$.

\subsection{Peak magnitudes}
It is convenient to analyze the evolution of the peak magnitudes with
$\Omega_{\s R}$
by contrasting it to the corresponding evolution in the absence of exchange. For $J=0$
there are two distinct regimes of the Rabi oscillations: weak driving, when $\Omega_{\s R}$ is smaller  than either $D$ or $\delta$, and strong driving,  $\Omega_{\s R}$ exceeds both
$D$ and $\delta$. In the first regime the Rabi spectrum is dominated by a  ``central" peak\cite{BoehmeMain} at $s=s_{\s a}, s_{\s b}$, see Eq. \eqref{sA},  when only one of the pair partners participates in the Rabi oscillations. In the second regime, the spectrum is dominated by peaks at ``large"
$s=s_{\s 2}=s_{\s a}+ s_{\s b}$ and ``small" $s=s_{\s 0}=|s_{\s a}-s_{\s b}|$. Most importantly,
the redistribution of intensities between the peaks happens monotonically\cite{Weakexchange} as $\Omega_{\s R}$
increases.

Compared to $J=0$, not only the positions of the peaks evolve with $\Omega_{\s R}$
 in a non-trivial fashion, but the redistribution of the peak intensities
with $\Omega_{\s R}$ is nonmonotonic. These intensities are represented by thicknesses of the spectral lines in Fig.~\ref{fig:FT}.
The magnitude of $s_{\s u}$-peak shown with red grows with $\Omega_{\s R}$. This
peak dominates the spectrum when its position approaches $s=2\Omega{\s R}$. The magnitudes of other two peaks, $s_{\s m}$ and $s_{\s l}$, vanish as they approach $s=\Omega_{\s R}$. However, the intensities
of these two peaks exhibit maxima for intermediate $\Omega_{\s R} \sim D$. Maximum in intensity of the $s_{\s l}$ peak, shown with yellow, is achieved when the position of this peak passes through a minimum.

In general, the intensities of all peaks turn to zero at small $\Omega_{\s R}$, which
is obvious on general grounds. A notable exception is the $s_{\s m}$-peak in Fig.~\ref{fig:FT}c, which corresponds to $\delta=\frac{3}{2}D$. It is seen from Eq. \eqref{asymptotic} that
for $\delta=\frac{3}{2}D$ the position of  the $s_{\s m}$-peak turns to $s=0$ in the limit
$\Omega_{\s R} \rightarrow 0$. For this reason the corresponding spectral line
is anomalously ``responsive" to a weak driving.

The evolution of the peak intensities with $\delta$ is also nonmonotonic, as illustrated in
Fig.~\ref{fig:FTdelta}. Naturally, the magnitudes of all peaks approach zero at
large enough detuning. The $s_{\s u}$-peak, which grew in intensity monotonically with $\Omega_{\s R}$, now falls off monotonically with $\delta$. Two other peaks, $s_{\s m}$ and $s_{\s l}$, again
have maxima at intermediate $\delta$. Maximum of the $s_{\s l}$-peak again corresponds
to the minimum in its position.

\subsection{Comparison to the simulation results Ref. \onlinecite{Simulations}}

A conventional way to study the Rabi spectra of coupled $S=1/2$ system adopted in Refs. \onlinecite{Boehme1,Boehme2,Boehme3} is based on direct numerical solution of the Liouville equation for $4\times 4$ density matrix. The quantity $\Delta\sigma(\tau)$ is then expressed through the diagonal
 elements of the density matrix and is, subsequently, Fourier transformed. The focus of Refs. \onlinecite{Boehme1,Boehme2,Boehme3} the effect
of detuning\cite{Boehme1}, exchange\cite{Boehme2}, and disorder\cite{Boehme3} caused by randomness of hyperfine field on the Rabi spectra. Dipole-dipole interaction was neglected in Refs. \onlinecite{Boehme1,Boehme2,Boehme3}. A comprehensive study which incorporates the competition between the
\begin{figure}[b]
\begin{center}
\includegraphics[width=78mm]{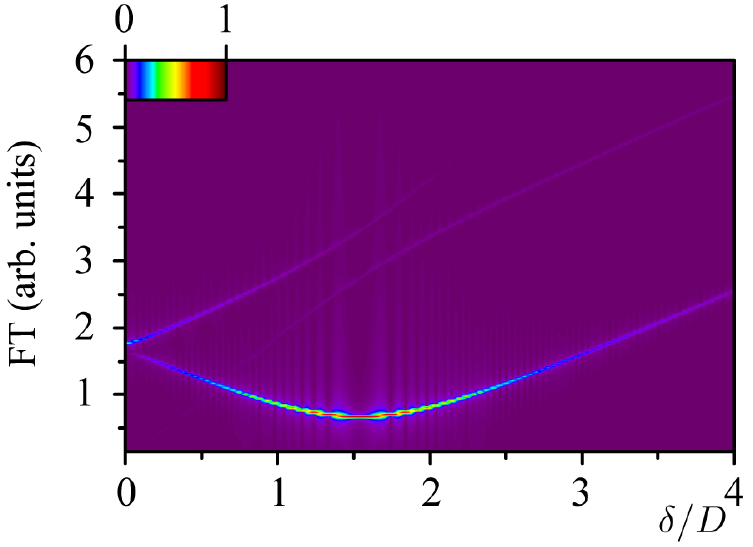}
\end{center}
\vspace{-0.3cm}
\caption{Spectrum of the Rabi oscillations obtained from numerical simulations for $\Omega_{\s R}/D=0.5$
as a function of dimensionless detuning $\delta/D$. The actual parameters used in simulations are $J=30$MHz,
$\Omega_{\s R}=1$MHz, $D=2$MHz. The line intensities are encoded in the brightness of the curves. }
\label{fig:simulation}
\end{figure}
exchange and dipole-dipole interactions was carried out recently in Ref. \onlinecite{Simulations}. It is
natural to compare the analytical results of the present paper obtained in the large-$J$ limit to the numerical results of Ref. \onlinecite{Simulations}. For this reason the simulations were run for the same
ratio $\Omega_{\s R}/D =0.5$ as in Fig.~\ref{fig:FTdelta}a. Similar to Fig.~\ref{fig:FTdelta}a the spectra were calculated versus
the dimensionless detuning $\delta/D$.
In simulations, the value of the exchange constant $J$ was chosen to be $30$MHz, which for chosen $D=2$MHz
ensures the large-$J$ limit, since $J/D = 15$.
The Rabi spectrum obtained from the numerical simulations is shown in Fig.~\ref{fig:simulation}. The intensities of the spectral lines are encoded in brightness of the curves. We see that the agreement of the analytical and numerical results is excellent: not only the positions of the lines agree perfectly with Fig.~\ref{fig:FTdelta}a, but also the  evolution of line intensities  with $\delta$ exhibits the same features,
the most prominent being the maximum of intensity at $\delta$ where the position of the line passes through a minimum.

\begin{figure}
\begin{center}
\includegraphics[width=80mm]{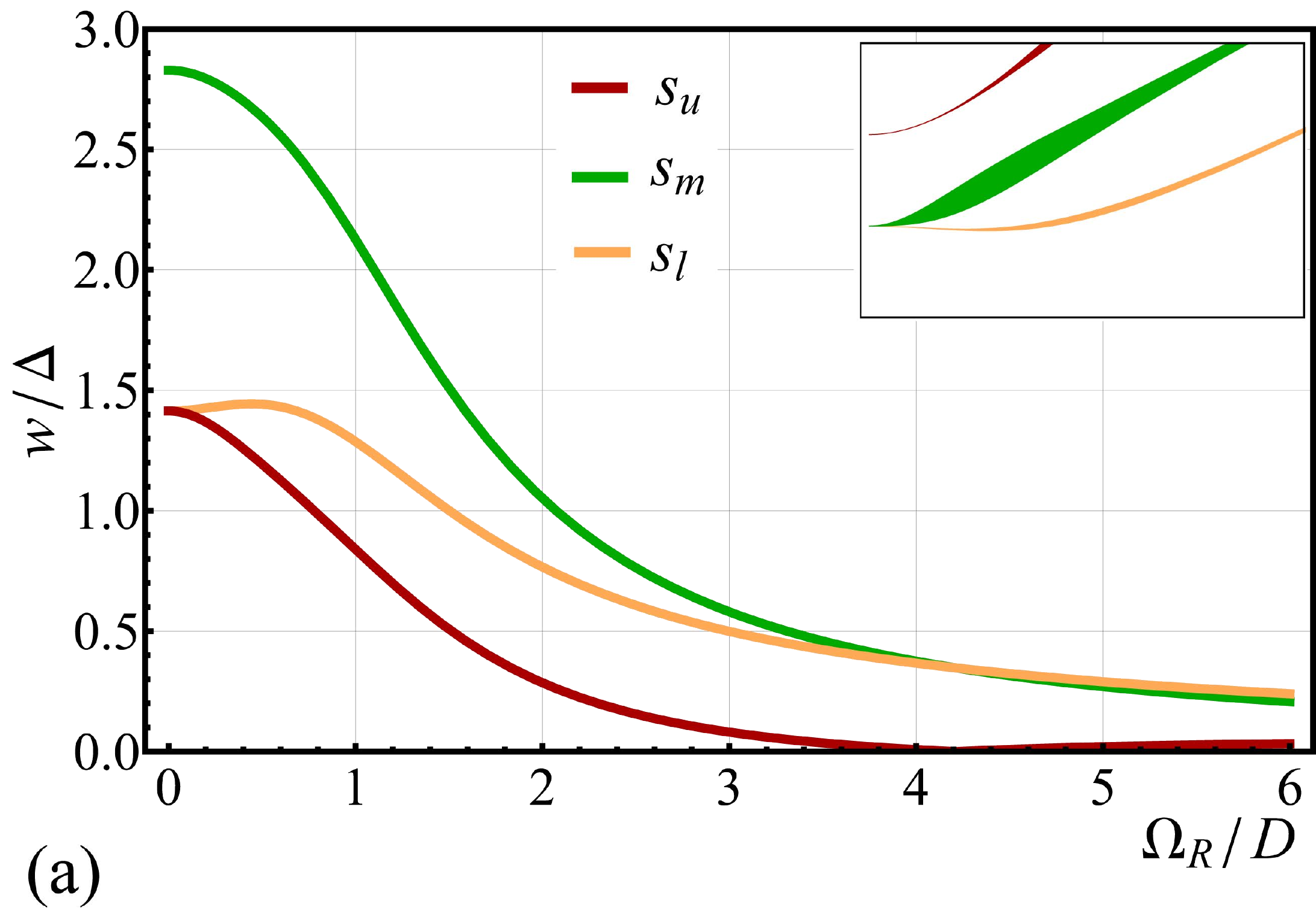}
\includegraphics[width=80mm]{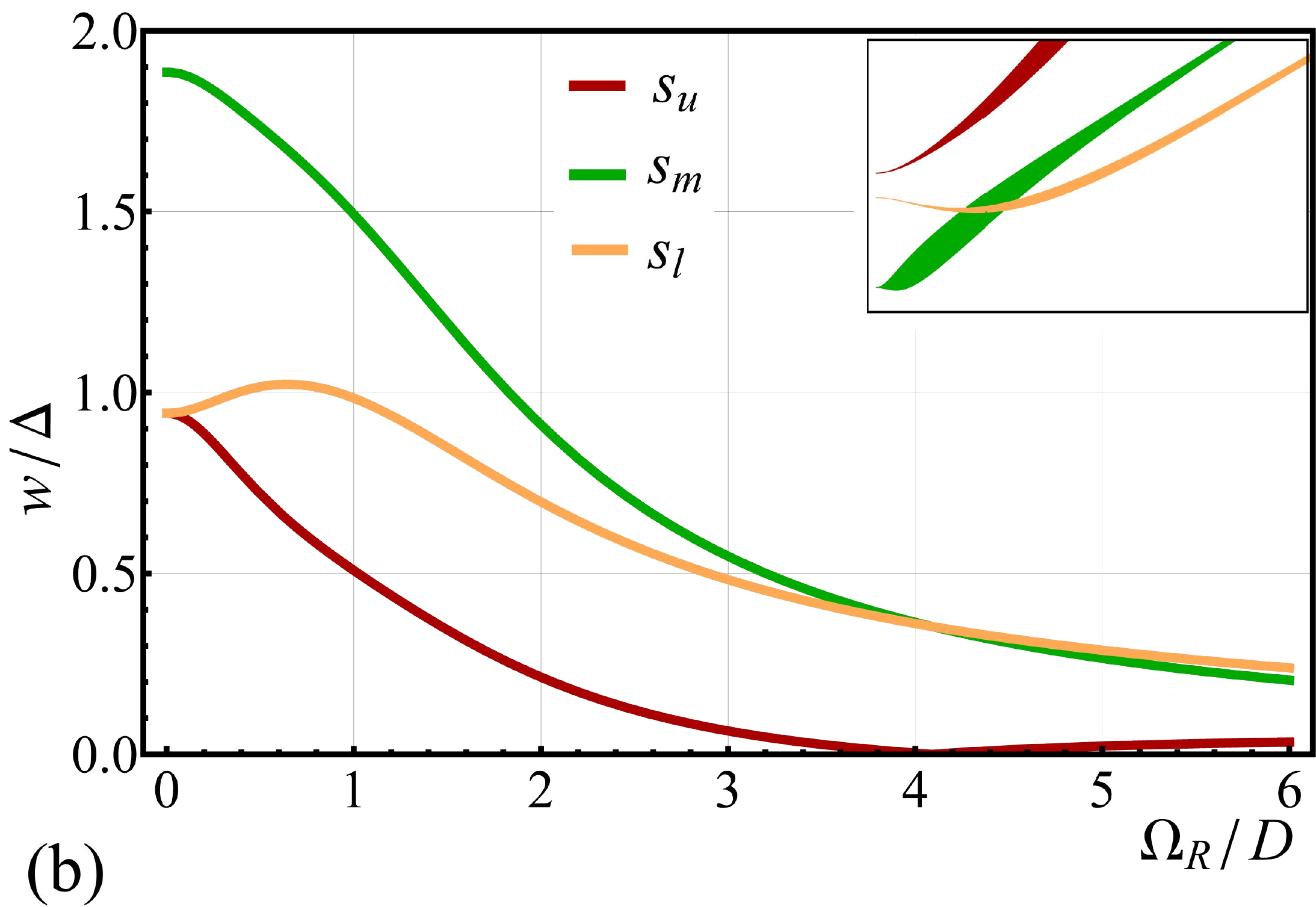}
\includegraphics[width=80mm]{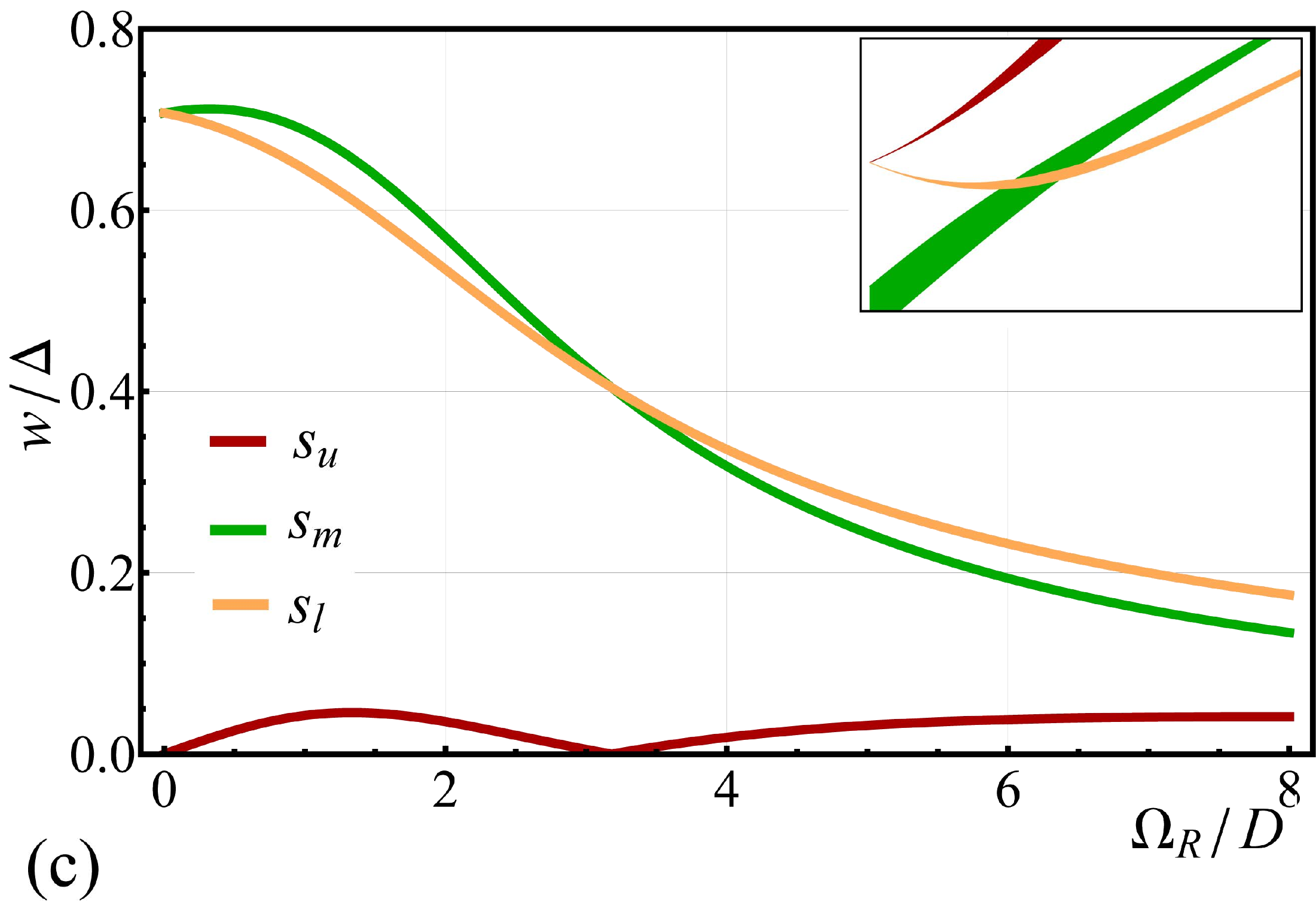}
\end{center}
\vspace{-0.7cm}
\caption{Broadening the Rabi spectral lines due to the spread, $\Delta$, in the local Larmour frequencies. Dimensionless widths, $w/\Delta$, of the Gaussian peaks are plotted from Eq. \eqref{correction} versesus $\Omega_{\s R}/D$, for three values of dimensionless detuning,  $\delta/D=0.5$ (a), $\delta/D=0.75$ (b), and $\delta/D=1.5$ (c). For convenience, the insets reproduce Figs.~\ref{fig:FT}a, b, c,  with corresponding positions of the centers of spectral lines.}
\label{fig:correction}
\end{figure}
\subsection{Peak widths}

Note that, unlike for Rabi oscillations in the absence of coupling,
the difference, $\delta_{\s 0}$,
between the Larmor frequencies of the pair partners (see Eq. (\ref{delta0})) does
not enter, neither into the positions of the spectral lines nor
into their intensities. The relevant quantity is the sum,
$\omega_{\s a}+\omega_{\s b}$, which enters into the detuning
parameter $\delta$, Eq. (\ref{detuning}).
In the ensemble of pairs the value of $\delta$ can fluctuate
from pair to pair due to,
 e.g., the randomness in nuclear
environment creating random hyperfine fields.
Weak disorder can be easily incorporated into the theory
since it transforms $\delta$-peaks
into Gaussians.

Suppose that Larmor frequencies, $\omega_{\s a}$ and $\omega_{\s b}$, are randomly
distributed around a central frequency, $\omega_{\s 0}$, with
width  $\Delta \ll \delta$.
To calculate the width, $w$, of each Gaussian we use  Eq. \eqref{cubic}.
Suppose that environment shifts detuning  from $\delta$ to $\delta+\delta_{\s 1}$, where
$\delta_{\s 1}=\frac{1}{2}(\omega_{\s a}+\omega_{\s b}-2\omega_{\s 0})$.
Then the quasienergy, $\tilde{\chi}_{\s i}$, acquires a shift $\delta_{\s 1}\frac{\partial \tilde{\chi}_{\s i}}{\partial \delta}$. This leads to three shifts of the peak positions of the form
$\delta_{\s 1}\left[\frac{\partial \tilde{\chi}_{\s i}}{\partial \delta}
-\frac{\partial \tilde{\chi}_{\s j}}{\partial \delta}\right]$.
Thus the width of the peak at $s=|\tilde{\chi}_{\s i}-\tilde{\chi}_{\s j}|$, resulting from quasienergies $i$ and $j$ is equal to
\begin{align}
\label{correction}
&w=\frac{\Delta}{\sqrt{2}}\left|\frac{\partial \tilde{\chi}_{\s i}}{\partial \delta}
-\frac{\partial \tilde{\chi}_{\s j}}{\partial \delta}\right|=
\sqrt{2}\Delta\Bigg[\frac{
\upsilon\delta(\eta_j-\eta_i)}{\frac{3}{4}(4\upsilon^2\eta_i^2-D^2)-(\delta^2+\Omega_{\s R}^2)}&
\nonumber \\
&\hspace{0.85cm}\times\frac{3(\upsilon\eta_i-D)(\upsilon\eta_j-D)-\frac{9}{4}D^2+\delta^2+\Omega_{\s R}^2}
{\frac{3}{4}(4\upsilon^2\eta_j^2-D^2)-(\delta^2+\Omega_{\s R}^2)}\Bigg].&
\end{align}
In the last identity we expressed $\frac{\partial \tilde{\chi}_{\s i}}{\partial \delta}$
through $\tilde{\chi}_{\s i}$ using Eq. \eqref{cubic}.

The widths of three peaks calculated from Eq. (\ref{correction}) are
plotted in Fig.~\ref{fig:correction}. It can be seen all three widths
fall off  with increasing $\Omega_{\s R}$. This is a quite natural behavior.
For convenience the insets in Fig.~\ref{fig:correction} show the positions
and intensities of the peaks in the absence of broadening. There is certain
correlation between intensities and the widths, namely, the middle peak, having
maximal intensity at small $\Omega_{\s R}$ has also the maximal width.
Overall, the ratio $\frac{w}{\Delta}$, characterizing how disorder, $\Delta$,
translates into the peak width lies within the range $0.2-1.2$.

\section{Concluding remarks}

We have derived an analytical description of spin--dependent electronic transition rates within strongly exchange coupled intermediate pairs during a coherent spin excitation which revealed the influence of several spin--Rabi oscillation components whose intensities and frequencies depend on the dipolar coupling within the pair, the strength of the driving field as well as the detuning. As long as the exchange coupling is strong, the oscillations controlling spin--dependent rates do neither depend of the value of the exchange coupling nor on the spin--orbit or hyperfine field induced Larmor separation (the difference of the pair partners Larmor frequencies). It must be emphasized that ``strong exchange" in the context of this work means that $J$ is larger than all other relevant parameters (dipolar strength, Larmor separation, driving field strength, and detuning) but not many orders of magnitude stronger. While the conclusions of the work presented here remain unchanged by the magnitude of $J$, very large $J$ will lead to rapidly decaying rate changes which will render the results irrelevant for their experimental application.

Obtaining the solutions analytically has been possible because in the limit of strong exchange, only three out of four spin states of
the pair participate in the Rabi oscillations. Note that there exists  another prominent object in which  Rabi oscillations
take place within a system of three levels. This object is a quantum dot
molecule\cite{Bayer,Finley,Gammon2,Gammon1,Gammon3,MoleculeRobust,MoleculeUlloa},
in which the excitation energy lies in the optical range.
The analog of spin is played by a two-level system consisting of
size-quantized electron and hole levels in a self-assembled quantum dot.
Experimentally, Rabi oscillations between these two levels are
studied by optical\cite{Sham,MacDonald},  and electrical\cite{Zrenner}  techniques.
The quantum dot molecule represents two vertically aligned quantum dots,
so that electron excited
in one dot can tunnel into another dot and vice versa. This tunneling
is a source of coupling between the dots which has no analog in spin
pair considered in the present paper. The other mechanism by which
different dots ``communicate" with each other is the Coulomb attraction
of excited electron to the hole left behind. If hole resides in one dot,
the energy of attraction of electron to this hole is different depending
whether electron resides in the same dot or in the neighboring dot.
This difference mimics the dipole-dipole interaction in the
spin system we considered. The most important difference between
the two systems lies in the structure of three levels participating in
the Rabi oscillations. In the spin pair these levels are shown in
Fig.~\ref{fig:spinstates}b and are almost evenly spaced in energy. In quantum dot
molecules the relevant levels are: the ground state with no exciton,
excited state with one exciton in the left dot, and excited state with
one exciton in the right dot, so that two excited states are close in
energy.

As a final remark, note that Rabi oscillations in quantum dot molecules
recently reported in Ref. \onlinecite{Gammon4}  correspond to simultaneous Rabi nutations
of {\em two} electrons. This is similar to the Rabi oscillations under
the conditions of half-field magnetic resonance which were also recently
observed experimentally\cite{BoehmeDifferentiation}.

\begin{acknowledgements}
We acknowledge D. P. Waters and R. Baarda for the preparation of the MEH-PPV diode pEDMR templates.
We also acknowledge the support of this work by the National Science Foundation through the Materials Research Science and Excellence Center (\#DMR-1121252). CB further acknowledges the support through a National Science Foundation CAREER award (\#0953225).
\end{acknowledgements}

\end{document}